\documentclass{amsart}
\makeatletter

\newtheorem{thm}{Theorem}[section]
\def\statetheorem{\@ifnextchar[{\@statetheorem}{\nr@statetheorem}}
\long\def\@statetheorem[#1]#2{\begin{thm}\label{#1}#2\end{thm}}
\long\def\nr@statetheorem#1{\begin{thm}#1\end{thm}}
\def\statetheorempf{\@ifnextchar[{\@statetheorempf}{\nr@statetheorempf}}
\long\def\@statetheorempf[#1]#2{\begin{thm}\label{#1}#2\end{thm}\proof}
\long\def\nr@statetheorempf#1{\begin{thm}#1\end{thm}\proof}

\newtheorem{lmma}{Lemma}[section]
\def\statelemma{\@ifnextchar[{\@statelemma}{\nr@statelemma}}
\long\def\@statelemma[#1]#2{\begin{lmma}\label{#1}#2\end{lmma}}
\long\def\nr@statelemma#1{\begin{lmma}#1\end{lmma}}
\def\statelemmapf{\@ifnextchar[{\@statelemmapf}{\nr@statelemmapf}}
\long\def\@statelemmapf[#1]#2{\begin{lmma}\label{#1}#2\end{lmma}\proof}
\long\def\nr@statelemmapf#1{\begin{lmma}#1\end{lmma}\proof}

\newtheorem{crlry}{Corollary}[section]
\def\statecorollary{\@ifnextchar[{\@statecorollary}{\nr@statecorollary}}
\long\def\@statecorollary[#1]#2{\begin{crlry}\label{#1}#2\end{crlry}}
\long\def\nr@statecorollary#1{\begin{crlry}#1\end{crlry}}
\def\statecorollarypf{\@ifnextchar[{\@statecorollarypf}{\nr@statecorollarypf}}
\long\def\@statecorollarypf[#1]#2{\begin{crlry}\label{#1}#2\end{crlry}\proof}
\long\def\nr@statecorollarypf#1{\begin{crlry}#1\end{crlry}\proof}

\def\ord{\textup{ord}\,}
\def\note{\@ifnextchar[{\@note}{\@note[Note]}}
\def\@note[#1]{\par\medskip\noindent{\textbf{#1:} }}
\def\C{\mathbb{C}}
\def\FF{\mathbb{F}}
\def\N{\mathbb{N}}
\def\bibref[#1]{\cite{#1}}

\let\real@bibitem\bibitem
\def\bibitem[#1]{\real@bibitem{#1}}

\makeatother

\title{Darboux Transformations of Bispectral Quantum Integrable Systems}

\author{Emil Horozov}
\address{Department of Mathematics and Informatics,
University of Sofia, Bulgaria}

\author{Alex Kasman}
\address{Department of Mathematics and Statistics, Concordia
University \\and\\ Centre de recherches math\'ematiques, Universit\'e de
Montr\'eal}
\curraddr{Mathematical Sciences Research Institute, Berkeley, CA}

\def\Spec{\hbox{Spec}\,}
\def\A{\mathcal{A}}
\def\ad{\textup{ad}}
\def\O{\mathcal{O}}
\def\D{\mathcal{D}}
\newcounter{examples}

\def\ord{\textup{ord}\,}
\begin{document}

\begin{abstract}{
We present an approach to higher dimensional Darboux transformations
suitable for application to quantum integrable systems and based on
the bispectral property of partial differential operators.
Specifically, working with the algebro-geometric definition of quantum
integrability, we utilize the bispectral duality of quantum
Hamiltonian systems to construct non-trivial Darboux transformations
between completely integrable quantum systems.  As an application, we
are able to construct new quantum integrable systems as the Darboux
transforms of trivial examples (such as symmetric products of one
dimensional systems) or by Darboux transformation of well-known
bispectral systems such as quantum Calogero-Moser.
}\end{abstract}
\maketitle

\section{Introduction}

Given three ordinary differential operators $L$, $K$ and $\tilde L$ we
say that $\tilde L$ is the {\it Darboux transform\/}
\cite{Darboux,Darboux2} of $L$ by $K$ if
they satisfy the intertwining relationship 
\begin{equation}
 K\circ L = \tilde L
\circ K.  \label{eqn:intertwining}
\end{equation}
In other words, Darboux transformation is nothing but the
conjugation of differential operators by differential operators.
Since this transformation preserves many spectral properties of $L$,
it has been extremely useful in the investigation of integrable
non-linear partial differential (i.e. soliton) equations (e.g.,
\cite{BHYsato,Kdt,Matveev,Matveev2}) and in the study of the bispectral
property (e.g., \cite{BHYbisp,DG,Kbisp}). Recently there has been much
interest in the generalization of Darboux transformations to higher
dimensional situations (e.g.,
\cite{Andrianov,BK,ChalVes,Kamran,Sabatier}) where it has similar
applications.

An ordinary differential operator $L(x,\partial_x)$ in the variable
$x$ is said to be {\it bispectral\/} \cite{DG,Gbisp} if it has a
family of eigenfunctions $\psi(x,z)$ $$ L\psi(x,z)=f(z)\psi(x,z) $$
parametrized by the spectral parameter $z$ (meaning that $f(z)$ is a
non-constant function) which is also an eigenfunction for an ordinary
differential operator $\Lambda(z,\partial_z)$ in $z$ with $x$ playing
the role of spectral parameter $$
\Lambda\psi(x,z)=\phi(x)\psi(x,z).
$$
A useful observation (cf. \cite{BHY}) is that the algebra $B$
generated by $L$ and $\phi(x)$ and the algebra $B'$ generated by
$\Lambda$ and $f(z)$ are related by an {\it anti-isomorphism\/}
\begin{eqnarray}
b:B&\to&B'\label{eqn:antiinva}\\
L&\mapsto&f(z)\\
\phi(x)&\mapsto&\Lambda\\
b(c+d) &=& b(c)+b(d)\\
b(cd) &=& b(d)b(c)\\
b(c) &=& 0 \ \hbox{iff}\ c=0\label{eqn:antiinvb}.
\end{eqnarray}

Although Darboux transformations have been frequently used to study the
bispectral property, this paper conversely will demonstrate the use of
bispectrality to construct Darboux transformations.  This is of
interest in the higher dimensional situation where determining
operators satisfying \eqref{eqn:intertwining} is quite difficult.  In
contrast, due to the simplicity of the factorization properties of
ordinary differential operators, it is easy to describe all possible
Darboux transformations in the one dimensional case:  Since one may
always factor an ordinary differential operator of order greater than
one, it is always possible to consider the general Darboux
transformation above as an iteration of Darboux transformations where
the dressing operator $K$ takes the form
\begin{equation}
K=f(x)\circ \partial\circ\frac{1}{f(x)}\label{eq:onestep}
\end{equation}
followed by conjugation by a function.  Thus, to determine all
equations \eqref{eqn:intertwining} for a particular ordinary
differential operator $L$ it is
sufficient to recognize that if $K$ is of the form
\eqref{eq:onestep} then there exists an ordinary differential
operator $\tilde L$  satisfying the equation if
and only if $f(x)$ is an eigenfunction of the operator $L$.

At present, there is no such general result in the higher dimensional
case.  Consequently, papers on the subject can only describe certain
classes of higher dimensional Darboux transformations and their
applications.  
Particularly
relevant to the present work is the paper \cite{BK} where, in analogy
to the form
\eqref{eq:onestep} from the one dimensional case, Darboux
transformations are considered for constant coefficient partial differential
operators by dressing operators of the form
\begin{equation}
K=p(x_1,\ldots,x_n)\circ
q(\partial_1,\ldots,\partial_n)\circ\frac{1}{p(x_1,\ldots,x_n)}\qquad p,q\in\C[x_1,\ldots,x_n]
\end{equation}
with $q$ irreducible.  

The present work generalizes that construction by presenting similar
results in the context of algebro-geometrically defined quantum
Hamiltonian systems \cite{BEG}.  After defining the notions of Darboux
transformations and bispectrality for quantum Hamiltonian systems we
demonstrate that the corresponding anti-isomorphism $b$
(cf.~\eqref{eqn:antiinva}-\eqref{eqn:antiinvb}) allows the
construction of non-trivial Darboux transformations
(Theorem~\ref{thm:big}).  As an application, new non-trivial examples
of quantum integrable systems can be constructed by Darboux
transformation of trivial examples and by Darboux transformation of
known bispectral examples such as the quantum Calogero-Moser system.

Although it is not necessary to the understanding of the present
paper, we would like to point out that the definition of Darboux
transformation for quantum integrable systems presented below is
inspired by the constructions in a number of papers
\cite{BHYsato,BHYbisp,BK,KR,Kdt,Liberati} which generalize and interpret the
results of G.~Wilson \cite{W} in terms of Darboux transformations of
rings of ordinary differential operators

\section{Quantum Integrable Systems}

Using the terminology of \cite{BEG}, we define a Quantum Hamiltonian
System (QHS) on a smooth, irreducible, affine algebraic variety $X$
(over the algebraically closed field $\FF$ in characteristic zero) to be a pair
$S=(\Lambda,L)$ where $\Lambda$ is also an affine variety over $\FF$ and
\begin{eqnarray}
L:\O(\Lambda)&\to& \D(X)\\
f&\mapsto& L_f
\end{eqnarray}
is an embedding of the coordinate ring of $\Lambda$ into the ring of
differential operators on $X$.

The system $S$ is said to be a Completely Integrable Quantum System
(CIQS) if $\dim \Lambda=\dim X$.  These systems are of particular
interest since they can be completely analyzed.
As described in \cite{BEG}, this is
the quantum analog of the algebro-geometric formulation of the
integrability of a classical Hamiltonian system.  In particular, one
should view the image of $L$ as commuting quantum Hamiltonians.

Associated to $S$ is a family of $\D$-modules parametrized by point
$\lambda\in\Lambda$ having generators $1_{\lambda}$ satisfying
$$
 L_{g}\cdot 1_{\lambda}=g(\lambda)1_{\lambda}\qquad \forall
g\in\O(\Lambda).
$$
One may think of this $\D$-module as a system of differential equations
$$
 L_{g}\cdot \psi=g(\lambda)\psi\qquad \forall
g\in\O(\Lambda)
$$
which is known as the spectral problem associated to $S$ \cite{BEG}.

\begin{examples}
\item\label{examp:weyl} The simplest example is $S_W^n$, which is the completely integrable system with
$\Lambda=X=\C^n$ given by $ L_{p}=p(\partial_1,\ldots,\partial_n)$ for
$p\in\C[x_1,\ldots,x_n]=\O(X)$ and
$\partial_i=\frac{\partial}{\partial x_i}$.  Note that $S_W^n$ is
actually nothing but the symmetric product of $S_W^1$ with itself $n$
times.

\item \label{examp:veselov}%
Let $\A$ be a finite set of vectors $\alpha\in\C^k$ and
$m_\alpha\in\N$ be a set of positive integers for $\alpha\in\A$.  Let
$L=-\Delta+u(x)$ ($x=(x_1,\ldots,x_k)$) be the Schr\"odinger operator where
$
\Delta=\partial_1^2+\partial_2^2+\cdots+\partial_k^2$ with
$\partial_j=\frac{\partial}{\partial x_j}$ and
$$
u(x)=\sum_{\alpha\in\A}\frac{m_\alpha(m_\alpha+1)(\alpha,\alpha)}{(\alpha,x)^2}.
$$
In \cite{bispV} and the literature cited therein one can find
necessary and
sufficient conditions on $\A$ and $m_{\alpha}$ such that
a wave function $\Psi(x,z)$ exists ($z=(z_1,\ldots,z_n)$) with the
property that $L\Psi=-||z||^2\Psi$.  Let $f$ be any polynomial
in $\C[x_1,\ldots,x_k]$ 
satisfying
$$
\partial_{\alpha}f=\partial^3_{\alpha}f=\cdots=\partial_{\alpha}^{2m_{\alpha}-1}f=0
$$
for any $x$ on the hyperplane $(\alpha,x)=0$ for each $\alpha\in\A$.
Then there corresponds a differential operator $L_f$ in $x$
such that $L_f\Psi=f(z)\Psi$.  Moreover, the coefficients of $L_f$ are
rational in $x$ having singularities only on the vanishing set of the
polynomal $p=\prod(x,\alpha)$.  So, in particular, one may choose any
sufficiently large 
finitely generated subring $A$ of such polynomials $f$ and consider
the completely integrable quantum system $S=(\Spec A,L)$ on the quasi-affine variety $X$
which is the localization of $\C^k$ by $p$.  
\end{examples}

\subsection{Localization}

Given a QHS $S=(\Lambda,L)$ on the variety $X$ and an embedding
$\gamma:\D(X)\to\D(Y)$ of the ring of differential operators on $X$ into
the ring of operators on some {\it other\/} variety $Y$, one obviously
can define a ``new'' quantum Hamiltonian system
$\hat S=(\Lambda,\gamma\circ L)$ on $Y$.  In general, one would not be
interested in doing so, since $\hat S$ is equivalent to $S$ in almost
every way.  However, a particular example of this construction will be
used in the following, and so it will be useful to develop an appropriate
notation here.

Let $X$ be an irreducible affine variety and $f\in\O(X)$ a non-zero
function in the coordinate ring.  The localization $\O_f$ constructed
from $\O(X)$ by formally introducing the inverse $\frac{1}{f}$ is the
coordinate ring of the open subvariety $X_f=\Spec(\O_f)$.  In
particular, $\dim X_f=\dim X$.
One may extend the action of $\D(X)$ to $\O(X_f)$ using the Leibniz
rule and thus view $\D(X)$ as a subring of $\D(X_f)$ embedded merely
by the inclusion $\gamma_f$.

\note[Definition] Given $S=(\Lambda,L)$ on $X$ and $f\not=0\in\O(X)$ define the
localization of $S$ by $f$ to be the QHS $S_f=(\Lambda,\gamma_f\circ
L)$ QHS on $X_f$. 

In order to simplify notation, we will make use of the embedding
$\gamma_f$ to identify $\D(X)$ with its image in $\D(X_f)$.  That is,
we will write $c\in\D(X_f)$ for any $c\in\D(X)$ rather than writing
$\gamma_f(c)$.

\section{Darboux Transformations for QHSs}

Let $S=(\Lambda,L)$ be a QHS on $X$. For any choice of
$K\in\D(X)$ let $A_K\subset \O(\Lambda)$ be defined as 
\begin{equation}
A_K=\{f\in\O(\Lambda)| K\circ  L_{f}=\tilde L_f \circ K \ \hbox{for some
$L_f\in\D(X)$}\}.  \label{eqn:A_K}
\end{equation}
In general, one would expect that $A_K=\FF$ (i.e., $A_K$ contains no
non-constant functions).  However, if
this is not the case, then $K$ may be used as a
``Darboux dressing'' to construct a new quantum system as follows.

Let $A\subset A_K$ be a finitely generated subring.  Then define the
 {\it Darboux transform\/} $\tilde S=DT(S,K,A)$ to be the pair
$\tilde S=(\tilde\Lambda,\tilde L)$ where
$\tilde\Lambda=\Spec A$ and $\tilde L_f$ is defined by equation
\eqref{eqn:A_K}. It is then elementary to verify that $\tilde S$ is
also a QHS on $X$.

Clearly, the difficulty then is in choosing an appropriate operator
$K$.  Although later in this paper we {\it will\/} describe below a
procedure for choosing a dressing operator $K$ leading to a
non-trivial Darboux transformation, thus far we have not given any
way in which to determine non-trivial Darboux transformations of QHSs
in dimension greater than one.  Below we will recall two examples from
earlier works which demonstrate Darboux transformations between
integrable systems.

\begin{examples}

\item \label{examp:CM3} It is shown in \cite{ChalVes} that certain
quantum Calogero models are Darboux transformations (according to the
definition above) of a localization of the quantum system $S_W^n$
(cf.~Example~\ref{examp:weyl}).  We will briefly recall here the
explicit computation in Section 3 of \cite{ChalVes} and ``translate''
it into the notation used above.  Let $S_p$ be the localization of
$S=S_W^3$ by the polynomial\footnote{Here we use the notation
$x_{ij}=x_i-x_j$ and $\partial_{ij}=\partial_i-\partial_j$.}
$p(x_1,x_2,x_3)=x_{12}x_{13}x_{23}$.  Then, one may consider the
Darboux transformation $\tilde S_p=DT(S_p,K,A)$ where
\begin{eqnarray*}
K &=& \partial_{12}\partial_{13}\partial_{23}-2x_{12}^{-1}\partial_{13}\partial_{23}-2x_{13}^{-1}\partial_{12}\partial_{23}-2x_{23}^{-1}\partial_{12}\partial_{13}\\&&+4x_{23}^{-1}x_{13}^{-1}\partial_{12}+4x_{13}^{-1}x_{12}^{-1}\partial_{23}+4x_{12}^{-1}x_{23}^{-1}\partial_{13}-12x^{-1}_{12}x^{-1}_{13}x^{-1}_{23}
\end{eqnarray*}
and where $A\subset\C[x_1,x_2,x_3]$ is the ring $\C[h_1,h_2,h_3]$ with $h_i=x^i_1+x^i_2+x^i_3$.  
Then $\tilde S$ is the three particle quantum
rational Calogero-Moser system.   In fact,
$\tilde L_{h_2}=\Delta-4\sum_{i<j} x_{ij}^{-2}$ is the standard Hamiltonian of
this well known integrable system.  (This is, in fact, a special case
of Example~\ref{examp:veselov}.)

\item\label{examp:bk} The first example in Section 5 of \cite{BK} is
a Darboux
transformation of a localization of $S=S_W^2$.  Let
$$K=\left(\partial_1-\frac{2x_1}{x_1^2-x_2}\right)\left(\partial_1+\frac{1}{x_1^2-x_2}\right)-\lambda
$$
$$\tau(x_1,x_2)=x_1^2-x_2\qquad
q(x_1,x_2)=x_1x_2-\lambda\in\C[x_1,x_2]=\O(\Lambda)$$
and $A=\C[q^3,x_1q^3,x_2q^3]$.
Then, it follows from the results of \cite{BK}
that for any fixed $\lambda\in\C$, the Darboux transform $DT(S_{\tau},K,A)$
is also a quantum integrable system.

\end{examples}

\section{Bispectrally Dual QHSs}

Consider a pair of quantum Hamiltonian systems (all varieties defined
over $\FF$):
$$
S=(\Lambda,L)\qquad\hbox{on}\ X
$$
$$
S'=(\Lambda',L')\qquad \hbox{on}\ X'.
$$
Suppose that $X'$ covers $\Lambda$ and that $X$ covers $\Lambda'$,
giving us natural embeddings
$$
\pi:\O(\Lambda)\to\O(X')\qquad\pi':\O(\Lambda')\to\O(X).
$$

Given $f\in\O(\Lambda)$ and $g\in\O(\Lambda')$ we have both $L_f\in\D(X)$
and $\pi'(g)\in\D(X)$.  
So, in particular, by using $L$ and $\pi'$ we can view
$\O(\Lambda)$ and $\O(\Lambda')$ as being subrings of $\D(X)$.
Similarly, $L'$ and $\pi$ allow us to consider both coordinate rings
as subrings of $D(X')$.

\note[Definition] Let $B\subset\D(X)$ be the subring generated by the images
of $L$ and $\pi'$.  Similarly, let
$B'\subset D(X')$ be the subring generated by the images of $L'$ and
$\pi$.  
We say that $S$ and $S'$ are {\it bispectrally
dual\/} to each other if there exists an anti-isomorphism $b:B\to B'$
such that
\begin{enumerate}
\item $b( L_{f})=\pi(f)\qquad \forall f\in\O(\Lambda)$
\item $b(\pi'(g))=L'_g\qquad\forall g\in\O(\Lambda')$
\item $b(L_1L_2)=b(L_2)b(L_1)$ for all $L_1,L_2\in B$
\item $b(L_1+L_2)=b(L_1)+b(L_2)$ for all $L_1,L_2\in B$
\item $b(L)=0$ iff $L=0\in B$.
\end{enumerate}

\note In order to simplify notation, we will use the embeddings $\pi$
and $\pi'$ to consider their images as actually being contained in the
coordinate rings of the covering varieties.  In particular, we will
write $g\in\O(X)\subset\D(X)$ for any $g\in\O(\Lambda')$ rather than $\pi'(g)$ and
similarly $f\in\O(X')\subset\D(X')$ for any $f\in\O(\Lambda)$.

The following lemma, whose statement is well known at least in the case of
bispectral ordinary differential operators \cite{DG,W} demonstrates
that the existence of the anti-isomorphism is a severe restriction.

\begin{lemma}[ad]{Denote by $A_i$ and $A_i'$ (for $i\in\N$) the
differential operators
$$
A_i=\ad_{ L_{f}}^i\left(g\right)\in\D(X)\qquad
A_i'=(-1)^i\ad_{f}^i\left(L'_g\right)\in\D(X').
$$
Then it follows from the bispectral duality that $b(A_i)=A_i'$ for all
$i\in\N$ and, in particular, that $A_i=0$ for $i>\ord L'_g$.}
When $i=0$ we simply have the known property $b(\pi'(g))=L'_g$
of the map $b$.  Then, supposing that the claim is known to apply for
$i$ we check that
\begin{eqnarray*}
b(A_{i+1})&=&b( L_{f}\circ A_i-A_i\circ  L_{f})\\
 &=& A_i'\circ \ f-\ f\circ A_i'\\
 &=& -\ad_{f}A_i'=A_{i+1}'.
\end{eqnarray*}
Since commutation with a function lowers
the order of an operator, it is clear that
$A_{i}'=0$ for any $i>\ord L'_g$.  However, the anti-isomorphism
demonstrates that $A_i=0$ for $i>\ord L'_g$ as well.
\end{lemma}

In general, for a differential operator $L$ of order greater than 0
and differential operator $g$ of order 0, one would expect that
$\ad_{L}^i g$ would be an operator of high order for arbitrarily large
$i\in\N$.  The fact that in the bispectral case the order is 0
demonstrates that bispectral duality is an unusual situation.
However, many well known examples of quantum integrable systems are
related by this duality.

\begin{examples}
As described in \cite{BHY}, whenever one has a bispectral situation --
commutative rings of operators in the variables $x$ and $z$
respectively sharing a common eigenfunction $\psi(x,z)$ with eigenvalues depending
on $z$ and $x$ respectively -- there automatically exists an
anti-isomorphism $b$ between the algebras generated by the operators
and eigenvalues in the separate variables.  Specifically, $b$ is the
map defined by the equation $L\psi(x,z)=b(L)\psi(x,z)$ for all $L\in B$.
In each of the examples
below, the existence of such a common eigenfunction is used to
demonstrate the bispectral duality of certain quantum integrable systems.
\item The differential operators from the system
$S_W^n$ are precisely the ring of constant coefficient differential
operators in the variables $x=(x_1,\ldots,x_n)$.  In particular, they
have the common eigenfunction $\Psi(x,z)=e^{(x,z)}$
($z=(z_1,\ldots,z_n)$).  Then, since this eigenfunction is shared also
by the constant coefficient operators in $z$, we may conclude
that $S_W^n$ is bispectrally dual to {\it itself\/}.
In particular, letting both $S$ and $S'$ be copies of $S_W^n$ one
finds that $B$ is the Weyl algebra and $b$ is the usual anti-isomorphism.
\item Similarly, since the eigenfunction $\Psi(x,z)$ from Example
\ref{examp:veselov} is symmetric in $x$ and $z$ \cite{bispV}, again
one finds that these systems are all bispectrally self-dual.
\item As described in
\cite{BK}, the CIQS $\tilde S_{\tau}$ on the local variety $\C^2_{\tau}$
from Example~\ref{examp:bk} is not-self dual, but is bispectrally dual
to a system $\tilde S'_{\tau}=(\tilde\Lambda',\tilde L')$ on
$\C^2_{x_1x_2-\lambda}$.

\item Since any commutative ring of bispectral differential
operators may be viewed as an example of a bispectral quantum
integrable system on a quasi-affine variety, one may find many
additional examples in the papers
\cite{BHY,BHYbisp,BK,KR,W}.
\end{examples}

\note[Remark]
Given completely integrable quantum systems  $S_i$ bispectrally dual
to the systems $S_i'$
($1\leq i \leq n$) the {\it symmetric products\/} are bispectrally
dual.  So, in particular, one may take {\it any\/} set of bispectral
ordinary differential operators (viewed as operators in different
variables) and this gives (trivial) bispectral quantum integrable
systems in any number of variables.  This is significant because the
techniques of the next section will allow us to transform these
trivial examples into non-trivial examples.

\section{Bispectral Darboux Transformations}

Let $S=(\Lambda,L)$ on $X$ and $S'=(\Lambda',L')$ on $X'$ be {\it
bispectrally dual completely integrable\/} quantum systems.  Let us
fix an arbitrary choice of $f\in\O(\Lambda)$ and $g\in\O(\Lambda')$.
Associated to this choice of a pair of functions we will construct by
Darboux transformation a new bispectrally dual pair of completely
integrable quantum systems: $\tilde S=(\tilde\Lambda,\tilde L)$ on the
localized variety\footnote{Recall that we are utilizing the embedding
$\pi'$ to consider $g(=\pi'(g))$ as an element of $\O(X)$.} $X_{g}$
and $\tilde S'=(\tilde
\Lambda',\tilde L')$ on $X'_f$.  Since the two Darboux dressing
operators to be used are related by the anti-isomorphism $b$, this is
an example of a {\it bispectral Darboux transformation}
\cite{BHY,KR,Kbisp}.

The key observation which will allow the Darboux transformations to be
performed are the following  factorizations which can be
achieved in the subring $B\subset\D(X)$:

\begin{lemma}[lem:factors]{Let $f$ and $g$ be as above, let $m=\ord
L_f$ and $n=\ord L'_g$, then there exist
elements $K,Q,R\in B$ satisfying
\begin{equation}
 g^{m+1}\circ L_f = K\circ g\label{eqn:factor1}
\end{equation}
\begin{equation}
L_f\circ g^{m+1} = g\circ R \label{eqn:factor2}
\end{equation}
\begin{equation}
L_f^{n+1} \circ g =  Q\circ L_f\label{eqn:factor3}
\end{equation}}

Denote by $I$ the left ideal $I=Bg\subset B$.  One has
$$
g\circ L_f=L_f\circ g+[g,L_f]
$$
i.e.
$$g\circ L_f=\ad_g L_f\ \textup{mod}\ I$$
where $\ad_g L_f=[g,L_f]\in B$ and has order at most $m-1$.

By the same argument, we can see that
$g^2\circ L_f=\ad_g^2 L_f\ \textup{mod}\ I$ where again $\ad_g^2 L_f\in
B$ and is an operator of order $m-2$ or less.  Continuing the
procedure one gets that 
$$
g^{m+1}\circ L_f=\ad_g^{m+1} L\ \textup{mod}\ I=0\ \textup{mod}\ I.
$$
Consequently we conclude that $g^{m+1}\circ L_f\in I$ which proves the
claim.

Equation \eqref{eqn:factor2}is proved in the same manner using instead the
 right ideal $gB$.

Moreover, by symmetry of definition we get equivalent factorization
in $B'$.  In particular, we see that $L'_g\circ f^{n+1}=f\circ R'$ for some $R'\in B'$.  Letting $Q=b(R')$ and applying the
anti-isomorphism $b$ to this equation yields \eqref{eqn:factor3}.
\end{lemma}

The main result then is that one may use the operator $K$ from
Lemma~\ref{lem:factors} as a Darboux dressing operator for the
localized system $S_g$ and one may use $b(K)$ as a Darboux dressing
operator for the localized system $S'_f$ and that the resulting
systems are also bispectrally dual quantum integrable systems.

\begin{theorem}[thm:big]{
\begin{enumerate}
\item[(a)]  Let $\{e_1,\ldots,e_N\}$ be a complete set of generators
of $\O(\Lambda)$ over $\FF$
and denote by $A$ the subring generated by $\{f^{n+1},f^{n+1}e_1,\ldots,f^{n+1}e_N\}$ where $n=\ord
L'_g$.  Then if $K$ is the operator defined in
\eqref{eqn:factor1} the Darboux transform $$\tilde S=(\tilde
\Lambda,\tilde L)=DT(S_g,K,A)$$ is a
completely integrable quantum system on $X_g$.

\item[(b)] Let $\{e'_1,\ldots,e_M'\}$ be a complete set of generators of
$\O(\Lambda')$ and denote by $A'$ the subring generated by $\{g^{m+1},e'_1
g^{m+1},\ldots,g^{m+1}e'_M\}$ where $m=\ord L_f$.  Then the Darboux
transform $$\tilde S'=(\tilde\Lambda',\tilde L')=DT(S'_f,b(K),A')$$ is a
completely integrable quantum system on $X'_f$.

\item[(c)] The systems $\tilde S$ and $\tilde S'$ are bispectrally dual.
\end{enumerate}}

(a) We must first show that $A$ is a subring of $A_K$
(cf.~\eqref{eqn:A_K}) so that the Darboux transformation is well
defined.  Note that by \eqref{eqn:factor3} we have that
$$
L_f^{n+1}=Q \circ L_f\circ g^{-1}\in \D(X_g).
$$
Moreover, from \eqref{eqn:factor1} we see that
$$
L_f\circ g^{-1}=g^{-m-1}\circ K\in\D(X_g).
$$
Combining these we conclude that
$$
L_f^{n+1}=Q\circ g^{-m-1}\circ K.
$$
Since every element of $A$
has a factor of $f^{n+1}$ (by definition), for any $h\in A$ we have
that $L_h=L_{h'}\circ L_f^{n+1}$ for some $h'\in\O(\Lambda)$.  Consequently
one has
$$
K \circ L_h =\tilde L_h\circ K
$$
with
\begin{equation}
\tilde L_h=K\circ L_{h'}\circ Q \circ g^{-m-1}.\label{eqn:tildeL}
\end{equation}
Thus, it follows that $\tilde S=(\tilde\Lambda,\tilde L)=DT(S_g,K,A)$
is a well defined QHS with $\tilde\Lambda=\Spec A$ and $\tilde L$
defined by \eqref{eqn:tildeL}.  

To see furthermore that this is an {\it integrable\/} system, it
suffices to note that the quotient field of the varieties $\Lambda$
and  $\tilde \Lambda$ are isomorphic.  Consequently, the varieties are
birational and their dimensions are equal.  (Essentially, we have
introduced singularities along $f^{-1}(0)\subset\Lambda$.)

\medskip 

\noindent (b) The proof of the second statement is the same, based on
the observation that applying the anti-isomorphism $b$ to
\eqref{eqn:factor1} gives 
$$
{L'_g}^{m+1}=f^{-1}\circ L'_g\circ b(K)\in\D(X'_f).
$$

\medskip

\noindent (c) We wish to demonstrate the existence of an
anti-isomorphism $\tilde b$ between the ring $\tilde B$ generated by
the operators $\tilde L_v$ and functions $w$ for $v\in \O(\tilde
\Lambda)$ and $w\in\O(\tilde\Lambda')$ and the ring $\tilde B'$ generated by
the operators $\tilde L'_w$ and functions $v$.  This follows
automatically from the existence of $b$ since we have conjugated by
$K$ and $b(K)$ respectively.  In particular, one may formally
(cf.~\cite{BHY}) define $\tilde b$ by its action on the generators:
$\tilde b(\tilde L_v)=b(K^{-1}\circ \tilde L_v \circ K)=b(L_v)=v$ and
$\tilde b(w)=b(K^{-1} w K)=b(K) L'_w b(K)^{-1}$.  Then the fact that it
extends to an anti-isomorphism on all of $\tilde B$ and $\tilde B'$
follows from the fact that it is just a conjugation of the
anti-isomorphism $b$.
\end{theorem}

\note[Remark] Since this procedure takes any bispectral quantum
integrable systems to a new pair of the same type, it may be iterated
indefinitely to produce new examples.

\begin{examples}
\item Let $\Lambda=X=\C^2$ and consider the CIQS $S=(\Lambda,L)$ given
by $L_p=p(\partial_1,\partial_2^2-x_2)$ for $p\in\C[x_1,x_2]$.  The
images of the generators $x_1$ and $x_2$ trivially commute since they
are simply ordinary differential operators in separate variables.  In
this way, we see that this is a trivial example since it is merely the
symmetric product of two one dimensional examples.  Moreover, since
the common eigenfunction $\psi(x,z)=e^{x_1z_1}Ai(x_2+z_2)$ satisfying
$L_p\psi=p\psi$ is symmetric in $x$ and $z$, this example is also
trivially self-dual.  (So, $b(p)=L_p$ and $b(L_p)=p$ defines the necessary
anti-isomorphism of the Weyl algebra in two variables.)  However, we
may use Darboux transformation to produce a new quantum system from
this one whose integrability and duality are not obvious.

Consider $f=x_1^2+x_2$ and $g=x_1+sx_2\in\C[x_1,x_2]=\O(\Lambda)$,
then $\ord L_f=\ord L_g=2$.  The results above allow us to consider
the Darboux transformation $\tilde S$ of the localization $S_g$ by the dressing
operator
\begin{eqnarray*}
K &=& g^{m+1}\circ L_f\circ
g^{-1}\\ &=& (x_1+sx_2)^2(\partial_{1}^2+\partial_2^2)-3(x_1+sx_2)s(\partial_1+s\partial_2)+[(x_1+sx_2)^2+2+2s^2].
\end{eqnarray*}
Specifically, in terms of the generators $e_1=1$, $e_2=x_1$, $e_3=x_2$
one finds that $\tilde S=(\tilde\Lambda,\tilde L)$ where
$\tilde\Lambda=\Spec \C[\tilde e_1,\tilde e_2,\tilde e_3]$ ($\tilde
e_i:=f^2e_i$) is a two dimensional
singular rational variety and $\tilde L$ is defined by its action on
the generators
$$
\tilde L_{\tilde e_i}=K\circ L_{e_i}\circ Q\circ g^{-3}
$$
with
$$
Q=L_f^2\circ(x_1+sx_2)+2L_f\circ(\partial_1+s\partial_2)+2.
$$

Moreover, one may also transform the system $S'$ localized at $f$ using the
dressing operator $b(K)$ to get the system $\tilde S'=(\tilde
\Lambda',\tilde L')$ where
$\tilde\Lambda'=\Spec \C[\tilde e_1',\tilde e_2',\tilde e_3']$
($\tilde e_i':=g^3e_i$) is also a two dimensional
singular rational variety and $\tilde L'$ is defined by its action on
the generators
$$
\tilde L'_{\tilde e_i'}=b(K)\circ L_{e_i}\circ f^{-1}\circ L_g.
$$

The fact that $\tilde S$ and $\tilde S'$ are bispectrally dual is
demonstrated by the fact that the eigenfunctions $\tilde \psi(x,z):=K
\psi(x,z)$ and $\tilde\psi'(x,z):=b(K)\psi(x,z)$ satisfying
$$ 
\tilde L_p\tilde\psi=p(z)\tilde\psi \qquad \tilde
L_q'\tilde\psi'=q(z)\tilde\psi'\qquad \forall p\in\O(\Lambda)\
q\in\O(\Lambda')$$ are related by the exchange of $x$ and $z$: $$
\tilde\psi(x,z)=\tilde\psi'(z,x).
$$

\item Of course, one may also use the construction
described above to determine completely integrable Darboux transforms of the quantum
Calogero-Moser system given in Example~\ref{examp:CM3}.  For example, it
now follows that one may use $K=h_i^3\circ \tilde L_{h_2}\circ h_i$ as a
Darboux dressing operator for this system to determine a new quantum
integrable system (having an operator of order $2i+2$ as its lowest order
Hamiltonian). 

\end{examples}

\note[Acknowledgements] Thanks to Bram Broer for helpful comments and
to Sasha Polishchuk for his advice and for referring us to the article
\cite{BEG}.  We are also grateful to John Harnad for his support
and assistance.  This work was partially supported by Grant MM-523/95
of the Bulgarian Ministry of Education, Science and Technology.


\begin{thebibliography}{10}
\bibitem[Andrianov]{A. Andrianov, N. Borisov and M. Ioffe, ``The
factorization method and quantum systems with equivalent energy
spectra'', Physics Letters A, 105 (1984) no 1,2 pp. 19--22.}

\bibitem[BHY]{B. Bakalov, E. Horozov, and M. Yakimov, ``General
Methods for Constructing Bispectral Operators'' {\it Physics Letters
A\/} 222 (1996) pp.~59--66}

\bibitem[BHYsato]{B. Bakalov, E. Horozov, M. Yakimov,
``B\"acklund-Darboux transformations in Sato's grassmannian, Serdica
Math. Jour. 22 (1996) 571-588.}

\bibitem[BHYbisp]{B. Bakalov, E. Horozov, M. Yakimov, ``Bispectral
commutative rings of ordinary differential operators'',
{\it Comm. Math. Phys.\/} 190 (1997) pp. 331--373}

\bibitem[BK]{Yu. Berest and A. Kasman, ``$\D$-modules and Darboux
transformations'', {\it
Letters in Mathematical Physics\/} 43 (1998) pp.~279--294}


\bibitem[BEG]{A. Braverman, P. Etingof and D. Gaitsgory, ``Quantum
Integrable Systems and Differential Galois Theory'', {\it
Transformation Groups\/} 2 (1997) pp. 31--56}


\bibitem[ChalVes]{O.A. Chalykh and A.P. Veselov, ``Commutative rings of
partial differential operators and Lie algebras'', Comm. Math. Phys.
125 (1990) 597-611.}

\bibitem[Darboux]{G. Darboux , ``{Sur une proposition relative aux
equation lineaires}'' Compt. Rend. {\bf 94} (1882), 1456}

\bibitem[Darboux2]{G. Darboux, ``Le\c{c}ons sur la th\'eorie
g\'en\'erale des surfaces.'' 2\'eme partie, Paris, Gauthiers-Villars,
1889.}

\bibitem[DG]{J.J. Duistermaat and F.A. Gr\"unbaum, ``Differential
equations in the spectral parameter'', Comm. Math. Phys. 103 (1986)
177-240}

\bibitem[Gbisp]{F.A. Gr\"unbaum, ``Bispectral Musings''
CRM Proceedings and Lecture Notes, Volume 14, American
Mathematical Society, Providence, RI, (1998) pp. 31--46.}

\bibitem[Kamran]{A. Gonzalez-Lopez and N. Kamran, ``The
Multidimensional Darboux Transformation'', {\tt hep-th/961200}.}

\bibitem[KR]{A. Kasman and M. Rothstein, ``Bispectral darboux
transformations: the generalized Airy case'', Physica D 102 (1997) p.
159-173.}

\bibitem[Kdt]{A. Kasman, ``Darboux transformations from $n$-KdV to
KP'', {\it Acta
Applicandae Mathematicae\/} 49 no.\ 2 (1997) pp.~179--197.}

\bibitem[Kbisp]{A. Kasman, ``Darboux transformations and the
bispectral problem'',  {\it CRM Proceedings and Lecture Notes\/}, 14, American Mathematical
Society, Providence, RI (1998) pp.~81--91.}

\bibitem[Liberati]{J.
Liberati,
``Bispectral property, Darboux transformation and the Grassmannian ${\rm Gr}^{\rm rat}$''
{\it Lett. Math. Phys.\/} 41 (1997), no. 4, pp. 321--332. }

\bibitem[Matveev]{V.B. Matveev,
   ``Darboux transformation and explicit solutions of the
   Kadomtcev-Petviaschvily Equation, Depending on
   Functional Parameters'' {\it  Letters in Mathematical Physics\/} 
   3 (1979), pp. 213-216}

\bibitem[Matveev2]{V.B. Matveev and M.A. Salle, {\it Darboux
Transformations and Solitons\/}, Springer-Verlag, Berlin, 1991.}

\bibitem[Sabatier]{P. Sabatier, ``On multidimensional Darboux
transformations'', Inverse Problems, 14 no. 2 (1998) pp. 355-366}


\bibitem[bispV]{A.P. Veselov, ``Baker-Akhiezer Functions and the
Bispectral Problem in Many Dimensions'', {\it CRM Proceedings and
Lecture Notes\/}  American
Mathematical Society, Providence, RI, 1998 pp. 123--127}

\bibitem[W]{G. Wilson, ``Bispectral commutative ordinary differential
operators'', J. Reine angew. Math. 442 (1993) 177-204}


\end{thebibliography}
\end{document}